\documentclass[12pt,preprint]{aastex}

\begin{document}

\title{Constrain intergalactic medium from the SZ effect map}

\author{Liang Cao\altaffilmark{1,2}, Wei Zuo\altaffilmark{1} and Yao-Quan Chu\altaffilmark{2,3}}

\altaffiltext{1}{Institute of Modern Physics, Chinese Academy of
Science, Lanzhou, Gansu 730000, China} \altaffiltext{2}{Center for
Astrophysics, University of Science and Technology of China,
Hefei, Anhui 230026, China} \altaffiltext{3}{National Astronomical
Observatories, Chinese Academy of Science, Chao-Yang District,
Beijing 100012, China}

\begin{abstract}

In this paper, we try to detect the SZ effect in the 2MASS DWT
clusters and less bound objects in order to constrain the warm-hot
intergalactic medium distribution on large scales by
cross-correlation analysis. The results of both observed WMAP and
mock SZ effect map indicate that the hot gas distributes from
inside as well as outside of the high density regions of galaxy
clusters, which is consistent with the results of both observation
and hydro simulation. Therefore, the DWT measurement of the
cross-correlation would be a powerful tool to probe the missing of
baryons in the Universe.

\end{abstract}


\section{Introduction}
The Sunyaev-Zeldovich(SZ) effect, or the inverse-Compton
scattering of the cosmic microwave background (CMB) photons by hot
electrons, shifts the spectrum of the CMB photons to higher energy
when the photons pass through the regions of cosmic hot gas.

Since galaxy clusters are major hosts of hot gas and hot
electrons, the SZ effect would lead to a correlation between the
maps of the CMB temperature fluctuations and the distribution of
galaxies on scales of clusters. Several groups have claimed the
detection of SZ effect signal with the cross-correlation between
first year WMAP data and samples of galaxy clusters.
Hern\'andez-Monteagudo \& Rubi\~no-Mart\'in(2004) performed a
cross-correlation test between 2MASS and WMAP for small patches on
the sky and found a strong detection; Afshordi et al.(2004)
detected the SZ effect at 3$\sigma$ level on small angular scales
by cross-correlating the WMAP data with the 2MASS galaxy
catalogue; Fosalba et al.(2004) reported a detection of the SZ
effect cross-correlating the WMAP data with large scale structure
traced by the SDSS data.

While Tukugita \& Peebles(2004) have argued that about 90\% of all
baryons are in the form of intergalactic plasma, but the
observational results on the properties of the intra-cluster
medium are quite uncertain, amount to $\approx 35\%$ of the total
are missing(Tukugita 2003). Since the SZ effect are significantly
indicated by cross-correlation,which makes the SZ detection a
powerful tool to constrain the hot gas in- and outside of clusters
of galaxies.

However, all these above detections of the SZ effect were found
mainly on small angular scales and mostly due to clusters.
Recently, there are also several authors focus on the
cross-correlation between WMAP and less bound objects such as
loose groups or super-clusters. Myers et al.(2004)
cross-correlated the WMAP data with the APM,ACO and 2MASS
catalogues and detect a SZ effect out to $\sim 1^{\circ}$ angular
distance of the cluster centers, which are interpreted as evidence
of cluster plasma as well as super-cluster hot gas. With the X-ray
based catalogues, Hern\'andez-Monteagudo et al.(2004) also detect
significant SZ effect from the clusters, but did not find SZ
evidence from super-clusters. These two discrepant results
motivate us care for the uncertainty to the model of gas
distribution, not simply assume that hot gas are traced by
galaxies.

We used the 2MASS extended source catalog and developed a biased
gas distribution model to study the SZ effect from clusters and
super-clusters. The SZ effects from 2MASS galaxies have been
unanimously detected(Myers et al. 2004; Afshordi et al. 2004),
which would be useful to test our results.The cross-correlation
statistics of this paper are based on the discrete wavelet
transform (DWT), which is more sensitive to detect the SZ
effect(Cao et al. 2006).

The paper is organized as follows. \S 2 describes the DWT
algorithm used in this paper. In \S 3,we perform the
cross-correlation analysis to constrain the hot gas in both
clusters and super-clusters through their SZ effect detection. The
discussion and conclusion are given in \S 4.

\section{Discrete wavelet transform(DWT) method}

\subsection{The DWT variables}
We describe the CMB temperature map and 2MASS galaxy distribution
by the variables using the 2-D orthogonal DWT decomposition
defined as Cao et al.(2006):
\begin{eqnarray}
\Delta T_{\bf j,l}=\frac{1 }
 {\int\phi_{\bf j,l}({\bf x})d{\bf x} }
  \int \Delta T({\bf x})\phi_{\bf j,l}({\bf x})d{\bf x}, \\
  \nonumber
\rho_{\bf j,l}=\frac{1 }
 {\int\phi_{\bf j,l}({\bf x})d{\bf x} }
  \int \rho_g({\bf x})\phi_{\bf j,l}({\bf x})d{\bf x},
\end{eqnarray}
and
\begin{eqnarray}
\tilde{\epsilon}^T_{\bf j,l} = \int \Delta T({\bf x})\psi_{\bf
j,l}({\bf x})d{\bf x}, \\ \nonumber \tilde{\epsilon}^g_{\bf j,l}=
\int \rho_g({\bf x}) \psi_{\bf j,l}({\bf x})d{\bf x},
\end{eqnarray}
where $\phi_{\bf j,l}({\bf x})$ and $\psi_{\bf j,l}({\bf x})$ are,
respectively, the scaling function and wavelet.The $\Delta T_{\bf
j,l}$ and $\rho_{\bf j,l}$ are called scaling function
coefficients (SFCs). For a $\Theta \times \Theta$ 2-D sample,they
describe, respectively, the mean temperature and the mean number
density of galaxies in the cell $({\bf j,l})$, which has size
$\Theta/2^{j_1} \times \Theta/2^{j_2}$ and at position around
$(l_1\Theta/2^{j_1}, l_2\Theta/2^{j_1})$, where $j_1$ and $j_2$
can be any integral, and $l_1=0,...2^{j_1-1}$,
$l_2=0,...2^{j_2-1}$. The angular distance between modes ${\bf l}$
and ${\bf l'}$ at scale $j$ is given by $\theta=\Theta\times|{\bf
l-l'}|/2^j$. The variables $\tilde{\epsilon}^T_{\bf j,l}$ and
$\tilde{\epsilon}^g_{\bf j,l}$ are called wavelet function
coefficients (WFCs) and describe, respectively, the fluctuations
of temperature and galaxy density on scale ${\bf j}$ at position
${\bf l}$ in wavelet space(Fang \& Feng 2000)..

\subsection{DWT clusters}
The SZ effect is sensitive to hot gas clouds.Since $\rho_{\bf
j,l}$ is proportional to the mean number density of galaxies, one
can identify the SZ effect signal with cross-correlation between
variables $\Delta T_{\bf j,l}$ and $\rho_{\bf j,l}$. In other
words, one can picked up top clusters on scale (size) ${\bf j}$ by
the top members of $\rho_{\bf j,l}$. These clusters are called DWT
clusters(Xu et al. 1999).

Using $\rho_{\bf j,l}$, we can identify clusters on various
scales. $j=8$ is on the angular scales of
$123^{\circ}.88/2^8\simeq 0^{\circ}.5$, which corresponds to
$\simeq 1.8$ h$^{-1}$ Mpc at the median redshift of the 2MASS
sample. Therefore, we will first identify the DWT clusters from
the 2MASS map by the top members of $\rho_{\bf j,l}$ on scales
$j=8$. For studying the SZ effect from less bound objects, we can
also pick up the DWT clusters on scale $j=7$ as super-clusters.

\section{SZ effect detection and intra-cluster hot gas distribution}

\subsection{Data samples}
We used the foreground cleaned WMAP maps of $W$ bands(Bennett et
al. 2003), from which the contamination of the galactic foreground
other than SZ effect is reduced. This sample is suitable to study
the SZ effect.While the galaxy sample are selected from the 2MASS
extended source catalog (XSC, Jarrett et al. 2000) with median
redshift $z \sim 0.1$.

To carry out the 2-D DWT analysis, we first take the equal-area
Lambert azimuthal projection,which projects the whole sky into two
circular plane.We then select a $\Theta \times \Theta$ square with
$\Theta = 123^{\circ}.88$ in the central part of each area.

\subsection{SZ effect of DWT clusters}
With these preparations, we can detect the temperature change of
the SZ effect with the cross-correlation between the WMAP data and
the 2MASS DWT clusters, defined by
\begin{equation}
\Delta T(|l-l'|)= \langle C_{\bf j,l}\Delta T_{\bf j,l'}\rangle,
\end{equation}
where the variable $C_{\bf j,l}$ is taken to be 1 for mode(cell)
(${\bf j,l}$) corresponding to a DWT cluster on scales $j$, and
$C_{\bf j,l}=0$, other modes. The average $\langle ...\rangle$
overs all possible $|l-l'|$ of the sample. Therefore, $\Delta
T(|l-l'|)$ is an average CMB temperature fluctuations with a
distance $|l-l'|$, or the corresponding angular distance
$123^{\circ}.88|l-l'|/2^j$ from the center of a DWT cluster.

The top panel of Fig. 1 presents the cross correlation $\Delta T$
vs.$\theta$ of $W$ band map with the top 500 $j=8$ DWT clusters.
The black points and error bars are, respectively, given by the
mean and 1-$\sigma$ of the ensemble $\Delta T(|l-l'|)$ of the
considered DWT cluster sample.

It shows an anti-correlation of the DWT clusters at $\theta=0$ or
$l=l'$ with temperature decrease $\Delta T_{sz}\equiv \Delta
T-\langle \Delta T \rangle \simeq -15 \pm 10$ $\mu$K. This result
is about the same as that given by 500 2MASS galaxy groups and
clusters selected by friends-of-friends algorithm (Myers et al.
2004). What's more,we can see that the anti-correlation occurs not
only at $\theta=0$, but also slightly at $\theta=0^{\circ}.5$.
Myers et al.(2004) also found that the strongest anti-correlation
is not only at the center of clusters, but on a angular distance
larger than a typical clusters. They suggests that it is due to
the SZ effect from the super-clusters hot gas.

To detect the SZ effect in less bound objects,we carry the similar
cross-correlation for 125 $j=7$ DWT clusters. We plots $\Delta T$
vs. $\theta$ for top 125 $j=7$ DWT clusters in the bottom panel of
Fig. 1. Within error bars,it shows no cross-correlation,i.e.,
there is no SZ effect due to super-clusters,which is consistent
with the results of Hern\'andez-Monteagudo et al.(2004).

Therefore, we conclude that the distribution of hot gas probably
is not simply proportional to the number density of optical and
infrared galaxies, the biased gas distribution model is needed.

\subsection{Biased gas distribution model}
With the assumption that galaxies trace hot baryon gas, i.e.
$n_e({\bf x})\propto \rho_g({\bf x})$, Cao et al.(2006) develop a
mock SZ effect map as
\begin{equation}
\Delta T^{Mock}_{sz}({\bf x}) = \Delta T_{cmb}-f \frac{\langle
(\tilde{\epsilon}^T_{\bf j,l})^2\rangle^{1/2}}
  {\langle (\tilde{\epsilon}^{g\alpha}_{\bf j,l})^2\rangle^{1/2}}
   \rho^{\alpha}_g({\bf x}),
\end{equation}
where $\tilde{\epsilon}^T_{\bf j,l}$ and
$\tilde{\epsilon}^{g\alpha}_{\bf j,l}$ are, respectively, the
wavelet variables of $\Delta T_{cmb}({\bf x})$ and
$\rho_g^{\alpha}({\bf x})$. The coefficient $f =\langle
(\tilde{\epsilon}^{T_{SZ}}_{\bf j,l})^2\rangle^{1/2}/
  \langle (\tilde{\epsilon}^{T}_{\bf j,l})^2\rangle^{1/2}$
is the ratio between the powers of SZ effect and CMB temperature
fluctuations on scale $j$, which is set to be 0.1 due to the
semi-analytical estimation(e.g. Cooray et al. 2004) in our
analysis.

In this paper,we improved the mock SZ effected CMB maps by the
follows: $\Delta T_{cmb}({\bf x})$ is given by the HEALPix
simulation, $\Delta T_{sz}({\bf x})$ is given by the second term
of eq.(4), but $\rho({\bf x})$ is taken to be the value given by
the 2MASS map if the position ${\bf x}$ is not only at the top 500
modes (${\bf j,l}$), but also around $|l-l|\leq 1$.

The top panel of Fig.2 plots the cross-correction between the mock
SZ effected CMB maps and 500 2MASS DWT clusters, the points and
error bars are respectively, the mean and 90\% range of 100 mock
samples. The mock sample surely yield the expected
anti-correlation $\Delta T$ decrease at $\theta=0$ and the tail at
$\theta \leq 0.^{\circ}5$. We also present the cross-correlation
between mock SZ effected CMB maps and 125 $j=7$ 2MASS
super-clusters in the bottom panel of Fig.2. Similar to the
results of WMAP observed map shown in Fig.1, the super-clusters SZ
effect disappeared,which supported again that the biased gas
distribution model are consistent with observations.

These results may be interpreted as that the tail of $\Delta T$ is
from the spread distribution of hot gas around the center of
clusters, but not from the super-clusters. The size of $j=7$ mode
is larger than $j=8$ mode by a factor of 4. However, top $j=7$ DWT
clusters from the 2MASS galaxies are less bounded systems, and
generally do not always contain top $j=8$ DWT clusters.Thus,as the
results of both observed and mock samples showed, there no SZ
signal can be seen with less bounded systems.

\section{Conclusion and discussion}

With the cross-correlation between WMAP and 2MASS, we have detect
the SZ effect in the 2MASS DWT clusters with temperature decrease
$\Delta T_{sz}\equiv \Delta T- \langle \Delta T \rangle \simeq -15
\pm 10$ $\mu$K, while there are no evidence of SZ effect in the
less bound objects. This result is consistent with the biased mock
SZ effected CMB map. Based on these results,we strongly suggests
that the hot gas are not traced by galaxy clusters,but more
outspread distributed.

As we have known, the galaxies formation and evolution involves a
complex dynamical system of separating baryonic gas and dark
matter,which is generally characterized by strong shocks. This
feature can be seen with the self-similar solution of spherical
collapse under the self-gravity of baryonic gas and dark matter
given by Bertschinger(1985). It shows that an outgoing shock is
always formed during the infall of baryons. The shocks outgoing
from high-density regions can slow down the infall motion of
baryons from low-density to collapsed regions,while dark matter is
not affected by the shocks. Thus, it can probably leads to the
decoupling between the mass and cosmic hot gas.

Our results that the distribution of the baryon fraction is not
uniform are consistent with both observations and simulations.The
X-ray measurements have revealed that the baryon fraction in
galaxy clusters is less than the prediction of primordial
nucleosynthesis (Ettori \& Fabian 1999). He et al.(2005) used the
hydro-simulation to study the spatial dependence cosmic baryons
distribution under the mechanism of shocks in nonlinear system,
they suggested that about 14\% baryons in the universe are
"hidden" around the clusters named warm-hot intergalactic
medium(WHIM).

On large scales, the non-uniformity of the baryon fraction is a
result of the statistical discrepancy of baryonic gas from the
underlying galaxy clusters.The baryons can be traced neither by
QSO absorption spectrum nor by X-ray emissions. However, the
ionized electrons in the outspread baryons would be scatter the
CMB photons and contribute to the SZ effect as well. Thus, the SZ
effect detection is a powerful tool to probe the missing of baryon
matter in the universe.

\acknowledgments

We greatly appreciate Prof. Li-Zhi Fang for his direction of this
study. We acknowledge the HEALPix software for producing
simulation CMB maps. The work was partly supported by National
Natural Science Foundation of China (10575119), Knowledge
Innovation Project of Chinese Academy of Science (KJCX2-SW-N02).

\clearpage

\begin{figure}
\figurenum{1}\epsscale{1.0}\plotone{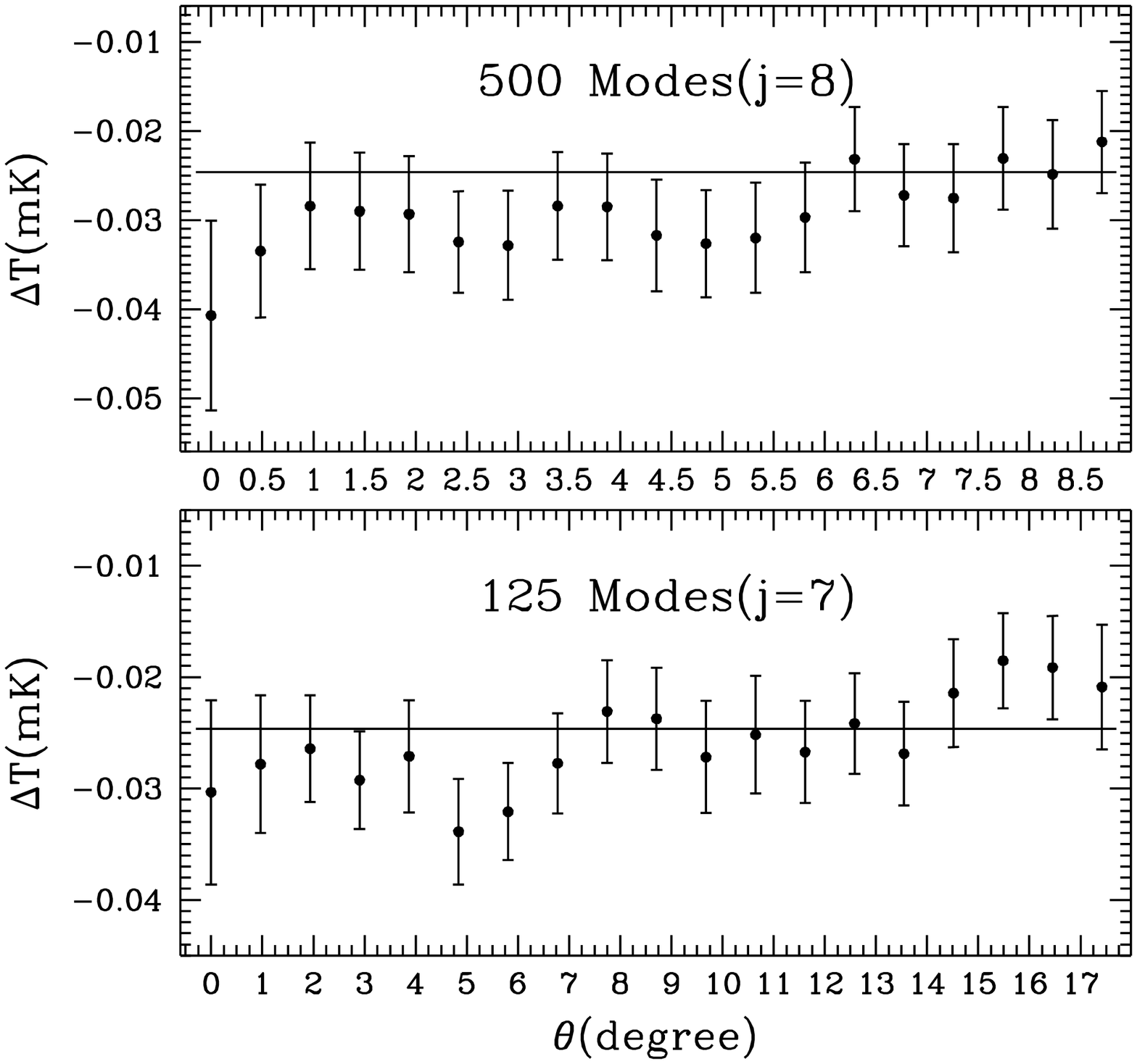}
\figcaption{Cross correlation $\langle T(|l-l'|)\rangle$
vs.$\theta$ between the WMAP map of $W$ band and 2MASS DWT
clusters.The top and bottom panels are, respectively, for top 500
$j=8$ and 125 $j=7$ DWT clusters.}
\end{figure}

\begin{figure}
\figurenum{2}\epsscale{1.0}\plotone{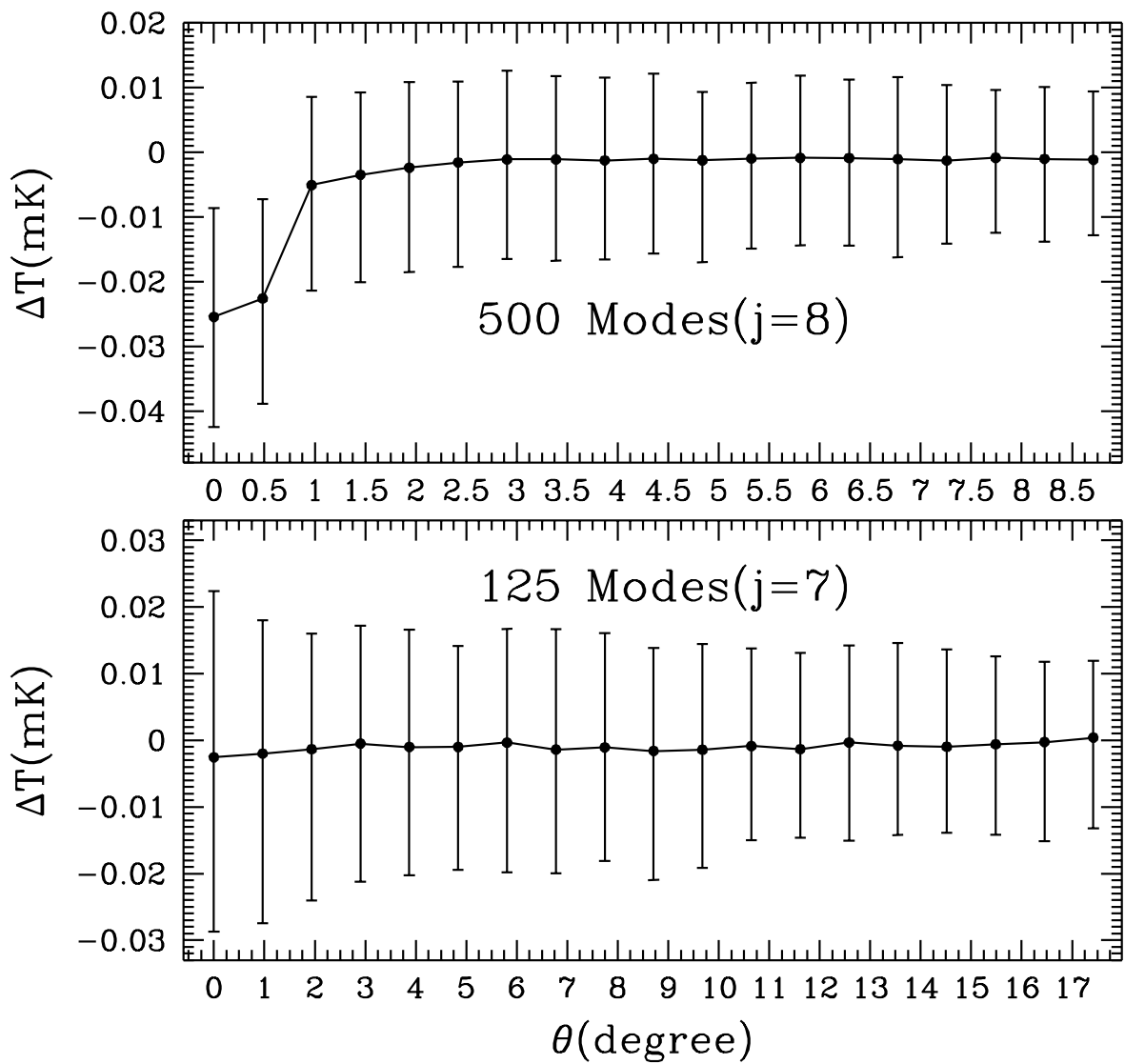}
\figcaption{Cross correlation $\langle T(|l-l'|)\rangle$
vs.$\theta$ between the Mock SZ effect map and 2MASS DWT clusters.
The top and bottom panels are, respectively, for top 500 $j=8$ and
125 $j=7$ DWT clusters.}
\end{figure}

\end{document}